# A Global Survey of Technological Resources and Datasets on COVID-19

Manoj Muniswamaiah, Tilak Agerwala, Charles C. Tappert

Seidenberg School of CSIS, Pace University, White Plains, New York

*Abstract*—**The application and successful utilization of technological resources in developing solutions to health, safety, and economic issues caused by COVID-19 indicate the importance of technology in curbing COVID-19. Also, the medical field has had to race against tie to develop and distribute the COVID-19 vaccine. This endeavour became successful with the vaccines created and approved in less than a year, a feat in medical history. Currently, much work is being done on data collection, where all significant factors impacting the disease are recorded. These factors include confirmed cases, death rates, vaccine rates, hospitalization data, and geographic regions affected by the pandemic. Continued research and use of technological resources are highly recommendable—the paper surveys list of packages, applications and datasets used to analyse COVID-19.**

*Keywords—Vaccination; hospitalization; confirmed cases; datasets; data science; COVID-19*

## I. INTRODUCTION

COVID-19 pandemic has affected the world; data is being collected by agencies, organizations, institutions, and other bodies that are keen on providing insights [391]. Data collection includes conducting case surveillance to gather data on demographics, clinical factors, epidemiologic characteristics, illness course, care, and history on exposure and contact. This data is needed to assess where, when, and who are most affected by the pandemic. The data available on COVID-19 is used by researchers in the medical field in evaluating different aspects of the virus. Statistical analysis tools estimate factors concerning the virus infectiousness obtained in growth rate and doubling time. Epidemiological models are used to group individuals based on their demographic data and apply mathematical formulas to find virus characteristics. Using machine learning, the time series prediction model is proposed to obtain the curve and forecast the epidemic's tendencies.

As a part of tackling the pandemic, different institutions, governments, organizations, and individuals have developed and adopted technological resources to manage the pandemic and support adherence to containment measures. Most of the technologies developed have utilized R programming, Python, Java, Kotlin, JavaScript, among other resources. Mobile applications were developed to help in contact tracing, notifications, and alerts to users if they interacted with a person infected. Dashboards have been used in visualizing COVID-19 cases across the world. This paper presents a survey of the technological resources and datasets been used by industry and academia to combat COVID-19.

Table I provide the name of the application; the details of the developer, institution, or academia; the summary of the application, its web link, and codebase link.

TABLE I.    TECHNOLOGICAL RESOURCES SUMMARY TABLE

| Application | Developer/industry/ Academia/University details | Application summary | Application weblink | Application code base link | References |
|---|---|---|---|---|---|
| Coronavirus tracker | Developed by John Coene. | Coronavirus tracker is an R Shiny app that tracks the spread of the coronavirus, based on three data sources, including John Hopkins, Weixin, and DXY Data. The app summarizes the coronavirus statistics such as deaths, confirmed, recovered, and suspected cases on a dashboard. | https://www.coronatracker.com/ [1] | https://github.com/JohnCoene/coronavirus.git [2] | [1] [2] [3] |
| COVID-19 Global Cases | Developed by Christoph Schoenenberger. | COVID-19 Global cases are a shiny app that displays the recent Covid-19 developments via key figures, plots, a map, and summary tables. | https://chschoenenberger.shinyapps.io/covid19_dashboard/ [4] | https://github.com/chschoenenberger/covid19_dashboard [5] | [3] [4] [5] |
| The 2019-20 Coronavirus Pandemic A timeline | Developed by Nico Hahn | Visualization of Covid-19 cases is a shiny application that uses leaflet, plotly, and data from Johns Hopkins University to visualize the novel coronavirus outbreak and show data for the entire world or particular countries. | https://nicohahn.shinyapps.io/covid19/ [6] | https://github.com/nicoFhahn/covid_shiny [7] | [3] [6] [7] |
| Modeling COVID-19 Spread vs Healthcare Capacity | Developed by Dr. Alison Hill from Johns Hopkins University | This application utilizes the epidemiological model based on the classic SEIR model to define the Covid-19 spread and clinical progression. The application provides different infection trajectories, clinical interventions to curb transmission, and a comparison to the current healthcare capacity. | https://alhill.shiny apps.io/COVID19 seir/ [8] | https://github.com/alsnhll/ SEIR_COVID19 [9] | [3] [8] [9] |
| COVID-19 Data Visualizatio n Platform | Developed by Shubhram Pandey. | This is a shiny app that provides an elaborate visualization of the impact of Covid-19 across the globe. The application also applies natural language processing from Twitter to provide sentiment analysis. | https://shubhramp andey.shinyapps.i o/coronaVirusViz/ [10] | https://github.com/shubhra mpandey/coronaVirus-dataViz [11] | [3] [10] [11] |
| Coronavirus 10-day forecast | This application was developed by Spatial Ecology and Evolution Lab (SpEEL) from the University of Melbourne. | It is a shiny app that provides a ten-day forecast of likely coronavirus cases by country, giving individuals a sense of how the Covid-19 is spreading or progressing. | https://covid19for ecast.science.unim elb.edu.au/ [12] | https://github.com/benflips /nCovForecast [13] | [3] [12][13] |
| Coronavirus (COVID-19) across the world | This application was developed by Anisa Dhana. | It is a shiny app that uses a map visualization of cases confirmed to monitor the spread of Covid-19 across the world and graphs to visualize the growth of the disease. | https://dash.datasc ienceplus.com/cov id19/ [14] | https://github.com/CSSEGI SandData/COVID-19 [15] | [16] [14] [15] |
| COVID-19 outbreak | Dr. Thibaut Fabacher developed this application in collaboration with the department of Public Health of the Strasbourg University Hospital and the Laboratory of Biostatistics and Medical Informatics of the Strasbourg Medicine Faculty. | The application displays an interactive map that indicates the worldwide monitoring of Covid-19 infection. The main area of focus of the app is on the evolution of the number of Covid-19 cases per country for a given period. | https://thibautfaba cher.shinyapps.io/ covid-19/ [17] | https://github.com/DrFabac h/Corona [18] | [3] [17] [18] |
| Corona trajectories | This application was developed by André Calero Valdez, from RWTH Aachen University. | The application uses two graphs to compare the number of confirmed cases and the deaths from Covid-19 with the country's trajectories. The application also allows users to compare the case number and growth rate of the Covid-19 pandemic per country using a table. | https://andrecalero valdez.shinyapps.i o/CovidTimeSerie sTest/ [19] | https://github.com/Sumidu/ covid19shiny [20] | [3] [19] [20] |
| Flatten the curve | Tinu Schneider developed | this application. In an interactive way, the app illustrates the different scenarios behind the #FlattenTheCurve message. | https://tinu.shinya pps.io/Flatten_the _Curve/ [21] | https://github.com/tinu-schneider/Flatten_the_Cur ve [22] | [3] [21] [22] |
| Explore the Spread of Covid-19 | Joachim Gassen developed this application, | The application allows users to visualize confirmed, recovered cases and reported deaths for several countries via one summary graph. | https://jgassen.shi nyapps.io/tidycovi d19/ [23] | https://statsandr.com/blog/t op-r-resources-on-covid-19-coronavirus/#tidycovid19 [24] | [3] [23] [24] |
| COVID-19 | Sebastian Engel-Wolf developed the application | The application visualizes elegantly collected Covid-19 data, including the confirmed cases, Maximum time of exponential growth in a row, deaths, populations, and Confirmed cases on 100,000 inhabitants, exponential growth, and the population. | https://sebastianw olf.shinyapps.io/C orona-Shiny/ [25] | https://github.com/zapping seb/coronashiny [26] | [3] [25] [26] |
| Simulation tool. COVID-19 epidemic in Togo - West Africa | Dr. Kankoé Sallah developed this application. | Uses SEIR metapopulation model with mobility between catchment areas to describe the country-level spread of COVID-19 and the impact of interventions in Togo, West Africa. | https://c2m-africa.shinyapps.i o/togo-covid-shiny/ [27] | | [3] [27] |
| Animating | Nathan Chaney | This application indicates a map animation of | | https://www.nathanchaney. | [30] [29] |
| COVID-19 hotspots over time | developed this application | new Covid-19 cases in the U.S.A measured in a 7-day rolling average. | | com/ [29] | |
| Covid-19-prediction | This application was developed by Manuel Oviedo and Manuel Febrero of Modestya research group of the University of Santiago de Compostela. | The application is a shiny app that provides a 5-day horizon prediction growth rate of Covid-19 using the evolution during the past 15-day growth rate. The prediction is performed using three functional regression models fitted and estimated on available data. Apart from the prediction values, the app provides an interactive table and plot for the expected number of accumulated cases and new daily confirmed and death cases. | http://modestya.us c.es:3838/covid19 prediction/ [33] | https://github.com/arnimpd m/Covid-19-prediction [34] | [3] [33] [34] |
| Healthcare worker deaths from novel Coronavirus (COVID-19) in the US | Jonathan Gross developed this application | The application is a shiny app that visualizes the U.S. health workers' deaths from Covid-19 reported on media outlets or news. The application is developed using R code with a map on the main page using Leaflet with tabs for additional graphs, including time series, histograms, and bar charts. | https://jontheepi.s hinyapps.io/hcwc oronavirus/ [35] | https://github.com/jontheep i/hcwcoronavirus [36] | [3] [35] [36] |
| Covid-19 Hospitalizations in Belgium | Jean-Michel Bodart developed this application | The dashboard indicates the hospitalizations related to Covid-19 in Belgium by province and region. | https://rpubs.com/ JMBodart/Covid1 9-hosp-be [37] | https://github.com/jmbo11 90/Covid19 [38] | [3] [37] [38] |
| Covidminder | This shiny app was developed by the Rensselaer Institute for Data Exploration and Applications | The application indicates the regional differences in determinants, medications, and outcome of the Covid-19 pandemic across the United but with a specific focus on New York. | https://covidminde r.idea.rpi.edu/ [39] | https://github.com/TheRen sselaerIDEA/COVIDMIN DER [40] | [41] [39] [40] |
| COVID-19 Canada Data Explorer | Petr Baranovskiy developed this application. | The application is a shiny app that analyses the official covid-19 dataset from the government of Canada and outputs the several indicators associated with the Covid-19 pandemic in the country. | https://dataenthusi ast.ca/apps/covid_ ca/ [42] | https://milano-r.github.io/erum2020-covidr-contest/petr-baranovskiy-covid-ca-data-explorer.html [43] | [3] [42][43] |
| P A G T A G N A: Philippine COVID-19 Case Forecasting Web Application | This application was developed by Jamal Kay Rogers and Yvonne Grace Arandela. | It is a shiny app that provides a 5-day forecast of Covid-19 cases in the Philipines include the confirmed new cases of infections, confirmed deaths, and recovery rate. Apart from forecasting, the application utilizes plots to visualize the disease's ten-day forecasts and the accumulated and confirmed data. The data used in this application is obtained from the Johns Hopkins University Center for Systems Science and Engineering (JHU CSSE). | https://jamalrogers app.shinyapps.io/sforecast/ [44] | https://github.com/fsmosca /COVID-19-PH-dataset [45] | [3] [44] [45] |
| COVID-19 Case & Death Report Number Corrector | Matt Maciejewski developed this application. | This shiny application is developed and aligned to make corrections of underreported Covid-19 cases and death. The application applies a multiplicative estimator for total deaths and cases regarding the base country to perform this role. | https://pharmhax.s hinyapps.io/covid-corrector-shiny/ [46] | https://github.com/pharmh ax/covid19-corrector [47] | [48] [46] [47] |
| COVID19 forecast | Carlos Catania developed this application | This application applies the SEIR model to forecast the spread of Covid 19 in various European and South American countries. | https://harpomaxx .shinyapps.io/covi d19/ [49] | https://github.com/harpom axx/COVID19 [50] | [3] [49] [50] |
| Trafford Covid-19 monitor | This is a shiny application developed by Trafford Data Lab | The application provides trends in confirmed coronavirus cases in Trafford. | https://trafforddata lab.shinyapps.io/tr afford_covid-19/ [54] | https://github.com/trafford DataLab/trafford_covid-19 [55] | [53] [54] [55] |
| Covid-19 Bulletin Board | Wei Su developed this application | The dashboard indicates the real-time Covid-19 visualization of the various covid-19 indicators in Japan, including the confirmed cases, hospital discharge and deaths, positive confirmed, and PCR test. | https://covid-2019.live/en/ [56] | https://github.com/swsoyee /2019-ncov-japan [57] | [3] [56] [57] |
| Covid-19 Statistics | Carl Sansfaçon developed this | It is a WordPress plugin that applies the R {ggplot2} graphics with ARIMA forecast and | http://moduloinfo. ca/wordpress/ [58] | https://plugins.trac.wordpre ss.org/browser/covid-19- | [3] [58] [59] |

| Displayer | application | PHP coding to display or visualize the confirmed new cases of Covid-19 infection, deaths, and recovered cases in various countries. The data used in this application is sourced from the COVID-19 Data Repository by the Centre for Systems Science and Engineering (CSSE) at Johns Hopkins University. | | statistics-displayer/ [59] | |
|---|---|---|---|---|---|
| CoronaMapper | This application was developed by Peter Gruber and Paolo Montemurro supported by OxyLabs. | The application visualizes the four-day average growth indicator of Covid-19 to indicate how the disease evolves after filtering out the noise. The visualizations are both interactive and intuitive. | http://coronamapper.com/ [60] | https://github.com/JayWelsh/coronamap [61] | [3] [60] [61] |
| CoronaDash | This is a shiny app developed by Peter Laurinec. | This application applies visualization and data mining techniques in R to compare Covid-19 statistics for different countries. The Covid-19 statistics displayed are obtained by using exponential smoothing model to extrapolate total confirmed cases; creating death trajectories; using dendrogram and table of clusters averages to create a multidimensional clustering; developing aggregated views of the entire world; and applying hierarchical clustering to compare the Covid-19 case between countries. | https://petolau.shinyapps.io/coronadash/ [62] | https://github.com/PetoLau/CoronaDash [63] | [3] [62] [63] |
| Covidfrance | This is a shiny app developed by Guillaume Pressiat | The application indicates the changes in the number of Covid-19 deaths and recoveries, hospitalization, and intensive care units by the department in France | https://guillaumepressiat.shinyapps.io/covidfrance/ [64] | https://gist.github.com/GuillaumePressiat/0e3658624e42f763e3e6a67df92bc6c5 [65] | [3] [64] [65] |
| COVID-19 Tracker | Dr Magda Bucholc developed this application from Ulster University | The application reports the number of reported Covid-19 cases at the local government district in Northern Ireland and county level across Ireland based on gender and growth rate. | https://nicovidtracker.org/ [66] | https://github.com/YouGov-Data/covid-19-tracker [67] | [68] [66] [67] |
| WHO COVID-19 Explorer | This application was developed by the World Health Organization (WHO) | This application provides timely updated data visualizations of Covid-19 cases, including confirmed cases and deaths by region and country. | https://worldhealthorg.shinyapps.io/covid [71]/ | https://github.com/WorldHealthOrganization/app [72] | [3] [71] [72] |
| COVID-19 Scenario Analysis Tool | The MRC Centre developed this application for Global Infectious Disease Analysis from the Imperial College London. | This application applies the squire R package to illustrate the Covid-19 pandemic trajectories, R_t & R_eff measures, and healthcare demand for different countries over time. | https://www.covidsim.org/v6.20210915/ [73] | https://github.com/mrc-ide/squire [74] | [3] [73] [74] |
| Coronavirus Package | Rami Krispin developed this R package | This package provides a clean dataset of the Covid-19 pandemic and analytics, including the daily summary of the pandemic cases by state. The dataset is collected from the John Hopkins database. | | https://ramikrispin.github.io/coronavirus/ [75] | [3] [75] |
| District Health Information Software (DHIS2) | The University of Oslo developed this application | The District Health Information Software has specific digital packages for Covid-19 that support the pandemic's surveillance and response activities. | https://www.dhis2.org/ [76] | https://github.com/dhis2/dhis2-covid19-doc [79] | [80] [76] [77] [78] [79] |
| Surveillance, Outbreak Response Management and Analysis System (SORMAS) | Helmholtz Centre for Infection Research developed this system. | The system performs the Covid-19 specific functions that are classified into aggregates and case-based functions. The aggregate functions include line listing, import, and export of data in CSV format; standard reporting of covid-19 cases including confirmed cases, deaths, and recoveries; and statistical analysis based on the reports provided by charts, maps, and graphs. The case-based functions include contact tracing, laboratory sample management, port of entry reporting, vaccination campaign, follow-up visit, and enrolling and tracing patients. | https://sormas.org/ [81] | https://github.com/hzi-braunschweig/SORMAS-Project [82] | [80] [81] [82] |

| Go.Data | This application was developed by WHO in collaboration with partners in the Global Outbreak Alert and Response Network (GOARN). | Since the outbreak of Covid-19 began, metadata packages have been prepared that match the most recent WHO Surveillance Guidance, including uniformity with all core metadata gathered as part of WHO Case Reporting Forms transmitted to COVID-MART / X-MART on a daily and weekly basis. If requested, this allows for streamlined IDSR reporting for countries. Other expanded metadata packages, such as the COVID First Few Hundred Cases (FFX) Protocol and the Unity Studies for HealthCare Workers, are available for countries conducting more extensive data collection or research inquiries. | https://www.who.int/godata [83] | https://github.com/godata-who/godata [84] | [80] [83] [84] |
|---|---|---|---|---|---|
| Epi Info | This application was developed by Centres for Disease Control and Prevention (CDC) | The Covid-19 specific functions include the development of COVID-19 Case Surveillance Forms that are customized for country, region, and local requirements. Epi Info is also applied in Covid-19 outbreak investigations, the development of small to mid-sized disease surveillance systems, the analysis, visualization, and reporting (AVR) components of larger systems, and continuing education in epidemiology and public health analytic methods at public health schools around the world. | https://www.cdc.gov/epiinfo/support/downloads.html [85] | https://github.com/Epi-Info/Epi-Info-Community-Edition [86] | [80] [85] [86] |
| Open Data Kit (ODK) | This application was developed by Get ODK, an organization majoring in data collection. | ODK software is being employed in the COVID-19 response for disease surveillance, fast diagnostics, and vaccine trials. | https://getodk.org/software/ [87] | https://github.com/getodk/collect [88] | [80] [87] [88] |
| CommCare | This software was developed by Dimagi, a firm providing digital data solutions | Dimagi created a set of pre-built COVID-19 template applications to help organizations and governments with their continuing COVID-19 response operations. | https://www.dimagi.com/covid-19/ [89] | https://github.com/dimagi/commcare-hq [90] | [80] [89] [90] |
| KoboToolbox | This software was developed by the Harvard Humanitarian Initiative, an organization working on the research and education of communities. | KoBoToolbox is a data collecting tool. | https://www.kobotoolbox.org/ [91][92] | https://github.com/kobotoolbox [93] | [80] [91] [92] [93] |
| Fast automated detection of COVID-19 from medical images | This application was developed by Shuang Liang & Huixiang Liu from School of Automation and Electrical Engineering, University of Science and Technology Beijing; Yu Gu from School of Automation, Guangdong University of Petrochemical Technology, Maoming; Xiuhua Guo, Zhiyuan Wu, Mengyang Liu & Lixin Tao<br><br>From the Department of Epidemiology and Health Statistics, | The software utilizes deep learning framework (neural network) that identifies COVID-19 from medical images. | | https://github.com/SHERLOCKLS/Detection-of-COVID-19-from-medical-images [94] | [95] [94] |

| | School of Public Health, Capital Medical University, Beijing, China; and Hongjun Li & Li from Beijing Youan Hospital, Capital Medical University, Beijing, China. | | | | |
|---|---|---|---|---|---|
| The Oxford Covid-19 Government Response Tracker (OxCGRT) | This application was developed by the Blavatnik School of Government. | The Oxford Covid-19 Government Response Tracker (OxCGRT) compiles systematic data on policy responses taken by countries to combat COVID-19. Since January 1, 2020, the various policy reactions have tracked over 180 nations and are categorized into 23 indicators, such as school closures, travel restrictions, and vaccination policies. These policies are scored on a scale to represent the magnitude of government intervention, and the results are compiled into a set of policy indices. The data can improve attempts to combat the epidemic by allowing decision-makers and citizens to understand government responses uniformly. | https://www.bsg.ox.ac.uk/research/research-projects/covid-19-government-response-tracker [96] | https://github.com/OxCGRT/covid-policy-tracker [97] | [98] [96][97] |
| COVID-19 Situazione Italia | This application was developed by the Department of Civil Protection (Dipartimento della Protezione Civile) Angelo Borrelli, Italy. | This application provides updated Covid-19 data and visualizations for Italy, including new confirmed infections, total confirmed infections, new confirmed deaths, total confirmed deaths, and recovered cases. The data and visualizations are provided for the whole country and the regions. | http://arcg.is/C1unv | https://github.com/pcm-dpc/COVID-19 [99] | [3] [99] |
| Covid Mobile data | This application was developed by COVID19 Mobility Task Force of the World Bank | The application uses the data from Mobile Network Operators (MNOs) to perform analytics. | | https://github.com/worldbank/covid-mobile-data [100] | [3] [100] |
| *Radar Covid-19* | The Government of Spain developed this application | This application was designed to prevent the spread of Covid-19. The application anonymizes users if they have had any contact in the last 14 days with someone infected with Covid-19 via low-power Bluetooth technology. | https://radarcovid.gob.es/ [101] | https://github.com/RadarCOVID/radar-covid-android [103] | [104] [101] [102] [103] |
| CovidSafe | The University of Washington developed this application. | The application was developed to help prevent the spread of Covid-19 by alerting users about highly relevant public health announcements, exposure to COVID-19 and to assist contact tracing without compromising users' privacy. | https://covidsafe.cs.washington.edu/ [105] | https://github.com/CovidSafe [106] | [3] [105] [106] |
| Covid Alert | Volunteers originally developed COVID Alert. The Canadian Digital Service is currently developing its repository. | This application was developed to slow down Covid-19 infections in Canada. The app notifies users if someone they were near in the past 14 days tells the app they tested positive. | https://www.canada.ca/en/public-health/services/diseases/coronavirus-disease-covid-19/covid-alert.html [107] | https://github.com/cds-snc/covid-alert-app [109] | [110] [107] [108] [109] |
| erouska-android | A team of volunteers initially developed this application. The application is currently developed and maintained by the Ministry of Health in collaboration with the National Agency for Communication | To combat the COVID-19 epidemic, the app alerts users at risk of spreading the virus. The software delivers guidance on how to minimize the spread of the epidemic based on the user's history of exposure to other potentially contagious users. | https://erouska.cz/ [111] | https://github.com/covid19cz/erouska-android [112] | [113] [111] [112] |
| | and Information Technologies (NAKIT) of the Czech Republic as part of the Smart Quarantine concept. | | | | |
| COVID-19 Dashboard | Johns Hopkins University Centre developed this application for Systems Science and Engineering. | This is the data repository for the Johns Hopkins University Centre for Systems Science and Engineering's 2019 Novel Coronavirus Visual Dashboard (JHU CSSE). The ESRI Living Atlas Team and the Johns Hopkins University Applied Physics Lab have also contributed to this project (JHU APL). | https://www.arcgis.com/apps/opsdashboard/index.html#/bda7594740fd40299423467b48e9ecf6 [114] | https://github.com/sidbannet/COVID-19_analysis [116] | [98] [114] [115] [116] |
| Corona-Warn-App | This application was developed as an open-source app by SAP and Deutsche Telekom under the directive by the government of Germany. | The Corona-Warn-App was developed with the goal of preventing the spread of Covid-19. The app serves as a digital complement to distancing, hygiene, and wearing masks. Additionally, it provides a functionality to add a user's digital vaccination certificate to prove their vaccination status. | https://www.coronawarn.app/en/ [117] | https://github.com/corona-warn-app/cwa-app-android [118] | [119] [117] [118] |
| TraceTogether | The Singapore Government Technology Agency developed this application. | Through community-driven contact tracing, TraceTogether supports Singapore's efforts to combat the spread of COVID-19. One can use the app to see or display their COVID Health Status based on their immunization and test results. | https://www.tracetogether.gov.sg/ [120] | https://github.com/OpenTrace-Community [121] | [122] [120] [121] |
| NZ COVID Tracer | The New Zealand Ministry of Health developed this application | The app helps contact tracing go faster by creating a private digital diary of the places you visit. Users Scan the official QR codes wherever they see them and add manual entries for their visits to other places. | https://www.health.govt.nz/our-work/diseases-and-conditions/covid-19-novel-coronavirus/covid-19-resources-and-tools/nz-covid-tracer-app [123] | https://github.com/minhealthnz/nzcovidtracer-app [125] | [126] [123] [124] [125] |
| VigilantGantry | This an automated contactless gantry system developed by GovTech's Data Science and Artificial Intelligence Division (DSAID) | VigilantGantry is an open-source implementation of an AI-driven automated temperature screening gantry that improves the rate of contactless screening by augmenting existing thermal systems. VigilantGantry is excellent for automatically scanning high-traffic sites for symptomatic COVID-19 patients. It helps ground crews keep on the lookout for COVID-19. | | https://github.com/dsaidgovsg/vigilantgantry [127] | [128] [127] |
| lancet-covid-19-database | Developed by Lancet | The Lancet COVID-19 Database gives users access to the most up-to-date information on COVID-19, such as cases, deaths, recoveries, testing, and other useful indicators for tracking the pandemic's spread and response. | | https://github.com/sdsna/lancet-covid-19-database [130] | [131] [130] |
| Covid19Canada | SDSN developed this application | This shiny app provides a forecast of Covid-19 cases and Covid-19 information in Canada, including the confirmed new cases of infections, confirmed deaths, and recovery rate. Apart from forecasting, the application utilizes plots to visualize the disease's ten-day forecasts and the accumulated and confirmed data. | https://art-bd.shinyapps.io/covid19canada/ [132] | https://github.com/ccodwg/Covid19Canada [133] | [3] [132] [133] |
| COVI-ML | This respiratory was developed by the Quebec Artificial Intelligence Institute | COVI-ML is the Risk model training code for the Covid-19 tracing application. Its repository provides models, infrastructure, and datasets for training deep-learning-based predictors of COVID-19 infectiousness as used in Proactive Contact Tracing. | | https://github.com/mila-iqia/COVI-ML [134] | [3][134] |

| Covid-19 model | Imperial College London developed this application/code. | This code was applied in modeling estimated deaths and infections for COVID-19 from the study "Estimating the effects of non-pharmaceutical interventions on COVID-19 in Europe "done by Flaxman et al. (2020) [136] | | https://github.com/Imperial CollegeLondon/covid19model [135] | [136] [135] |
|---|---|---|---|---|---|
| The COVID Tracking Project | Alexis Madrigal created this project through a collaborative volunteer-run effort to track the ongoing COVID-19 pandemic | This project collects and publishes data required to understand the COVID-19 outbreak in the United States. All 50 states, five territories, and the District of Columbia participate in the Covid tracking project, which will collect data on COVID-19 testing and patient outcomes. | https://covidtracking.com/ [137] | https://github.com/COVID 19Tracking [138] | [139] [137] [138] |
| Covidmx | Covidmx was developed by Federico Garza | The API was developed to handle Covid-19 open data provided by the Mexican Dirección General de Epidemiología. | | https://github.com/Federic oGarza/covidmx [140] | [141] [140] |
| Covid19-Scenarios | Neherlab developed this tool | The Covid-19 Scenarios provide Models of generating trajectories for COVID-19 outbreak and hospital demand. The functioning of this tool is based on the SIR model, which simulates a COVID19 outbreak. | https://covid19-scenarios.org/ [142] | https://github.com/neherlab /covid19_scenarios [143] | [3] [142] [143] |
| covid-chest-imaging-database | This database was developed by NHSX and the British Society of Thoracic Imaging (BSTI). NHSX is a joint unit of National Health Service (NHS) England and the Department of Health and Social Care, supporting local NHS and care organizations. | The database was developed with a respiratory containing tooling related to the NHSX National COVID-19 Chest Image Database (NCCID) to promote research projects in response to the COVID-19 pandemic. | | https://github.com/nhsx/co vid-chest-imaging-database [144] | [145] [144] |
| Covid-pass-verifier | This is an application developed by NHSX | The COVID Pass Verifier app is the official NHS COVID Pass Verifier for England and Wales. The app is a safe and secure way to check if someone has been appropriately vaccinated against COVID-19, has had a negative test, or has recovered from COVID-19. | https://www.nhsx. nhs.uk/covid-19-response/nhs-covid-pass-verifier-app/international-covid-pass-verifier-app-user-guide/ [146] | https://github.com/nhsx/co vid-pass-verifier [148] | [145] [146] [147][148] |
| Covasim | The Institute for Disease Modelling developed this simulator | Covasim is a stochastic agent-based simulator for performing COVID-19 analyses. | | https://github.com/Institute forDiseaseModeling/covasi m [149] | [150] [149] |
| covid-19 Dashboard | Greg Rafferty developed Covid-19 dashboard | This is a web dashboard developed to monitor the COVID-19 pandemic. The data used is obtained from Johns Hopkins Center for Systems Science and Engineering. | https://covid-19-raffg.herokuapp.c om/ [151] | https://github.com/raffg/co vid-19 [152] | [3] [152] [151] |
| Covid-19 R/Python scripts | Developed by QuKunLa; a Laboratory of Immunogenomics and Precision Medicine, University of Science and Technology of China | These are R/Python scripts to analyze single-cell RNA-sequence data from COVID-19 patients. | | https://github.com/QuKun Lab/COVID-19 [153] | [3] [153] |
| COVID-19-CT-CXR | COVID-19-CT-CXR was developed by Peng et al. and Intramural Research Programs of the | This is a public database of COVID-19 CXR and CT images, which are automatically extracted from COVID-19-relevant articles from the PubMed Central Open Access (PMC-OA) | | https://github.com/ncbi-nlp/COVID-19-CT-CXR [154] | [155] [154] |

| | National Institutes of Health, National Library of Medicine and Clinical Centre. | Subset. | | | |
|---|---|---|---|---|---|
| covid19-healthsystemcapacity | This project was developed by the CovidCareMap organization | This application assists in better understanding, anticipating, and acting to support and ramp up our health systems' capacity (beds, staffing, ventilators, supplies) to effectively care for a rapidly growing number of active COVID19 patients in need of hospitalization and intensive (ICU) care. | | https://github.com/covidcaremap/covid19-healthsystemcapacity [156] | [3] [156] |
| CV19 Index | The Global Loop team developed this model | The COVID-19 Vulnerability Index (CV19 Index) is a predictive model that identifies persons who are more susceptible to COVID-19 severe problems. The CV19 Index is designed to assist hospitals, federal, state, and local public health agencies, and other healthcare organizations in identifying, planning for, responding to, and reducing COVID-19's impact in their areas. | https://www.closedloop.ai/covid-19-index [157] | https://github.com/closedloop-ai/cv19index [158] | [159] [157] [158] |
| OpenABM-Covid19 | This model was developed by the Pathogen Dynamics Group of Oxford Big Data Institute. | OpenABM-Covid19 is an agent-based model (ABM) that was created to model the spread of Covid-19 in a city and investigate the impact of passive and active intervention measures. | | https://github.com/BDI-pathogens/OpenABM-Covid19 [160] | [3] [160] |
| COVID-19 vaccination slot booking script | PythonRepo developed this script. | Is used to automate covid vaccination booking. | https://pythonrepo.com/repo/pallupz-covid-vaccine-booking [161] | https://pythonrepo.com/repo/pallupz-covid-vaccine-booking [162] | [159] [161] [162] |

TABLE II.        COVID-19 DATASETS SUMMARY TABLE

| Developer/Industry/Academia/ University/Organization Details | Dataset Summary | Dataset Usage | Weblink to the Dataset | References |
|---|---|---|---|---|
| Our World in Data | Data on COVID-19 vaccinations that include country-by country statistics of the COVID-19 vaccines administered to date. | Vaccine outreach program. | https://ourworldindata.org/covid-vaccinations. [164] | [163][164] |
| | Data on COVID-19 confirmed deaths per country. | Effects of testing, managing, hospitalization. | https://ourworldindata.org/covid-deaths.[165] | [362][165][394] |
| | Global data of confirmed COVID-19 cases | Effect on travel restriction, intervention programs. | https://ourworldindata.org/covid-cases. [166] | [393][166] |
| | Data on COVID-19 testing, i.e., positivity rate, contact tracing, tests performed per day | Pandemic preventive measures | https://ourworldindata.org/coronavirus-testing. [167] | [393][167] |
| | Data on COVID-19 hospitalization | Monitoring cases to improve impact on available resources. | https://ourworldindata.org/covid-hospitalizations.[168] | [392][168] |
| | COVID-19 mortality risks | Segregation of age groups that may be at risk of dying from the disease. | https://ourworldindata.org/mortality-risk-covid. [169] | [393][169][170] |
| | Excess mortality due to COVID-19 | Segregation of age groups that may be at risk of dying from the disease, and | https://ourworldindata.org/excess-mortality-covid. [171] | [393][171] |
| | | other accelerating factors. | | |
| | Policy responses to the COVID-19 pandemic | Government interventions to curb the spread of the virus. | https://ourworldindata.org/policy-responses-covid. [172] | [393][172][173] |
| The Johns Hopkins University Center for Systems Science and Engineering [JHU CCSE] | COVID-19 Epidemiological Data | For segmentation of COVID-19 cases based on epidemiological characteristics. | https://data.humdata.org/dataset/novel-coronavirus-2019-ncov-cases. [174] | [174] |
| OCHA | COVID-19 number of confirmed cases, deaths, and recoveries by the province in Indonesia | Mobility transmission analysis. | https://data.humdata.org/dataset/indonesia-covid-19-cases-recoveries-and-deaths-per-province.[175] | [175] |
| World Health Organization | COVID-19 cases and deaths | Mobility transmission and mortality analysis. | https://data.humdata.org/dataset/coronavirus-covid-19-cases-and-deaths.[176] | [176] |
| Blavatnik School of Government, University of Oxford | OXFORD COVID-19 Government Response Stringency index | Government measures | https://data.humdata.org/dataset/oxford-covid-19-government-response-tracker. [177] | [177] |
| HDX | COVID-19 Vaccinations | Rate of vaccine drives | https://data.humdata.org/dataset/covid-19-vaccinations. [178] | [178] |
| World Food Program | COVID-19 global airline information and travel restriction | Global Monitoring. | https://data.humdata.org/dataset/covid-19-global-travel-restrictions-and-airline-information. [179] | [179] |
| HDX | COVID-19 cases and deaths in the United States | Reporting cases at a national level. | https://data.humdata.org/dataset/nyt-covid-19-data. [180] | [180] |
| HDX | Total number of COVID-19 tests performed per country | Monitoring cases | https://data.humdata.org/dataset/total-covid-19-tests-performed-by-country. [181] | [181] |
| UNESCO | Global school closures | Area segmentation | https://data.humdata.org/dataset/global-school-closures-covid19. [182] | [182] |
| Meta | FAIR COVID-19 US County Forecast | Country-level forecast. | https://data.humdata.org/dataset/fair-covid-dataset.[183] | [183] |
| CARE Bangladesh | District Wise Quarantine for COVID-19 | Reporting cases at a national level. | https://data.humdata.org/dataset/district-wise-quarantine-for-covid-19. [184] | [184] |
| HDX | COVID-19 Impact on Humanitarian Operations Data Viz inputs | Reporting humanitarian activities at a national level. | https://data.humdata.org/dataset/covid-19-data-visual-inputs. [185] | [185] |
| OCHA Venezuela | COVID-19 sub-national data | Reporting cases at a national level. | https://data.humdata.org/dataset/corona-virus-covid-19-cases-and-deaths-in-venezuela.[186] | [186] |
| OCHA FISS | Global Humanitarian Operational Presence Who, What, Where [3W] Portal | Reporting humanitarian activities at a global level. | https://data.humdata.org/dataset/ocha-global-humanitarian-operational-presence-who-what-where-3w-portal.[187] | [187] |
| ACAPS | COVID-19 Government Measures Dataset | Reporting government measures at a global level. | https://data.humdata.org/dataset/acaps-covid19-government-measures-dataset.[188] | [188] |
| HDX | Europe COVID-19 subnational cases | COVID-19 infected area segmentation. | https://data.humdata.org/dataset/europe-covid-19-subnational-cases. [190] | [190] |
| OCHA Philippines | Philippines COVID-19 response. | Reporting government measures at a national level. | https://data.humdata.org/dataset/philippines-covid-19-response-who-does-what-where. [191] | [191] |

| Code for Venezuela | COVID-19 education impact survey | Monitoring impact on a national level | https://data.humdata.org/dataset/open_one_time_covid_education_impact. [192] | [192] |
|---|---|---|---|---|
| iMMAP | Google mobility report | Mobility transmission analysis. | https://data.humdata.org/dataset/google-mobility-report. [193] | [193] |
| Humanitarian Emergency Report Africa [HERA] | Subnational data on Covid 19 cases per day | COVID-19 infected area segmentation. | https://data.humdata.org/dataset/nigeria_covid19_subnational.[194] | [194] |
| HDX | Worldwide geographic distribution of COVID-19 cases | COVID-19 infected area segmentation. | https://data.humdata.org/dataset/ecdc-covid-19. [195] | [195] |
| World Health Organization | Immunization campaigns impacted due to COVID-19. | Mobility transmission analysis | https://data.humdata.org/dataset/immunization-campaigns-impacted. [196] | [196] |
| HDX | Excess mortality during COVID-19 pandemic | Segregation of age groups that may be at risk of dying from the disease. | https://data.humdata.org/dataset/financial-times-excess-mortality-during-covid-19-pandemic-data. [197] | [197] |
| HDX | COVID-19 subnational cases in Palestine | Reporting cases at a national level. | https://data.humdata.org/dataset/state-of-palestine-coronavirus-covid-19-subnational-cases.[198] | [198] |
| Meta | Impact survey and trends on COVID-19 | Reporting cases at a national level. | https://data.humdata.org/dataset/covid-19-symptom-map. [199] | [199] |
| HDX | COVID-19 vaccine doses are given to humanitarian resource plan countries | Forecasts on dose availability and actual deliveries | https://data.humdata.org/dataset/covid-19-vaccine-doses-in-hrp-countries. [200] | [200] |
| World Bank Group | World Bank indicators of interest to the COVID-19 outbreak | Data for use in response, modeling analysis | https://data.humdata.org/dataset/world-bank-indicators-of-interest-to-the-covid-19-outbreak.[201] | [201] |
| Global Health 50/50 | Gender and COVID-19 project | Exploring how gender may be driving the higher proportion of reported deaths in men among confirmed cases so far. | http://globalhealth5050.org/covid19[202] | [202] |
| World Bank Group | Harmonized data on Household COVID-19 monitoring surveys | Data analysis and trend checking | https://data.humdata.org/dataset/harmonized-covid-19-household-monitoring-surveys[203] | [203] |
| Humanitarian Emergency Response Africa [HERA] | African continent Covid 19 cases | Data analysis and trendsetting | https://data.humdata.org/dataset/covid19_africa_continental-infections-recoveries-deaths[204] | [204] |
| Dalberg | Developing countries' government action on COVID-19 | non-pharmaceutical interventions | https://data.humdata.org/dataset/government-actions-on-covid-19[205] | [205] |
| Meta | Survey on preventative health | Monitor and understand people's knowledge and practices about COVID-19 to improve communications and their response to the pandemic. | https://data.humdata.org/dataset/preventive-health-survey[206] | [206] |
| International Organization for Migration | Information on populations within the Far North region of Cameroon | Providing regular, accurate, and updated data to better support the response of the Government of Cameroon and the humanitarian | https://data.humdata.org/dataset/cameroon-baseline-assessment-data-iom-dtm[207] | [207] |

| | | community. | | |
|---|---|---|---|---|
| HDX | Covax round 6 allocations | Monitoring of Covax vaccine allocations | https://data.humdata.org/dataset/covax-round-6-allocations.[208] | [208] |
| Humanitarian Emergency Response Africa | COVID-19 subnational data in Burkina Faso | Reporting Covid data at National level | https://data.humdata.org/dataset/burkinafaso_covid19_subnational[209] | [209] |
| Metabiota | Spatiotemporal data for COVID-19 deaths and cases. | Data analysis and monitoring | https://data.humdata.org/dataset/2019-novel-coronavirus-cases[210] | [210] |
| HDX | COVID-19 subnational data for Afghanistan | Data analysis and reporting on a national level | https://data.humdata.org/dataset/afghanistan-covid-19-statistics-per-province[211] | [211] |
| Cuebiq Inc | COVID-19 mobility data for Italy | Monitoring mobility changes in Italy since lockdown | https://data.humdata.org/dataset/covid-19-mobility-italy[212] | [212] |
| Humanitarian Emergency Response Africa | COVID-19 subnational cases in Africa | Reporting COVID-19 cases on a national level | https://data.humdata.org/dataset/africa-coronavirus-covid-19-subnational-cases[213] | [213] |
| Qatar Computing Research Institute | Twitter data geographic distribution of COVID-19 | Geographical distribution of twitter users and tweets regarding COVID-19 pandemic. | https://data.humdata.org/dataset/covid-19-twitter-data-geographic-distribution[214] | [214] |
| ACAPS | Secondary impacts of Covid 19 on a global scale | Aid Decision-making on addressing wider effects of COVID-19 | https://data.humdata.org/dataset/global-covid-19-secondary-impacts.[215] | [215] |
| Humanitarian Emergency Response Africa | COVID-19 city level in Burkina Faso | Reporting Covid data at a city level | https://data.humdata.org/dataset/burkinafaso_covid19_city-level[216] | [216] |
| Hub Latin America | The COVID-19 mortality rate in Lima, Peru | Reporting, analysis of COVID-19 death rates in Lima | https://data.humdata.org/dataset/peru-covid19-mortality-rate-in-lima[217] | [217] |
| Infoculture | COVID-19 cases in Moscow | Statistics | https://data.humdata.org/dataset/covid-19-cases-data-in-moscow[218] | [218] |
| HDX | Social measures and public health applied during COVID-19 | Analysis and reporting. | https://data.humdata.org/dataset/world-global-database-of-public-health-and-social-measures-applied-during-the-covid-19-pandemic[219] | [219] |
| Mobile Accord, Inc [GeoPoll] | Impact and perceptions of Coronavirus in Sub-Saharan African countries | Analysis and reporting | https://data.humdata.org/dataset/covid-19-impacts-africa[220] | [220] |
| HDX | Subnational COVID-19 cases for Iraq | Reporting Covid data at National level | https://data.humdata.org/dataset/iraq-coronavirus-covid-19-subnational-cases[221] | [221] |
| HDX | Covid 19 related funding from IATI | Monitoring of funding use in fighting COVID-19 | https://data.humdata.org/dataset/iati-covid19-funding[222] | [222] |
| HDX | Gavi and World Bank COVID-19 vaccine funding | Fund disbursement and support for COVID-19 | https://data.humdata.org/dataset/world-bank-and-gavi-vaccine-financing[223] | [223] |
| Code for Venezuela | Survey on COVID-19 impact | Data analysis and interpretation | https://data.humdata.org/dataset/open_one_time_covid_impact[224] | [224] |
| Humanitarian Emergency Response Africa | COVID-19 cases in Ethiopia | Reporting cases at a national level. | https://data.humdata.org/dataset/ethiopia-covid19-cases[225] | [225] |
| World Bank Group | High frequency indicators for COVID-19 | Data analysis and interpretation | https://data.humdata.org/dataset/covid-19-high-frequency-indicators[226] | [226] |

| OCHA FISS | Global humanitarian response plan COVID-19 administrative boundaries and population-statistics | Reporting cases at a national level. | https://data.humdata.org/dataset/global-humanitarian-response-plan-covid-19-administrative-boundaries-and-population-statistics[227] | [227] |
|---|---|---|---|---|
| INFORM | Inform Risk Index for COVID-19, Version 0.1.4 | Support prioritization of preparedness and early response actions for the direct impacts of the pandemic and identify countries where secondary effects are likely to have the most critical humanitarian consequences. | https://data.humdata.org/dataset/inform-covid-19-risk-index-version-0-1-4[228] | [228] |
| Safeture | COVID-19 subnational cases in Kazakhstan | For data analysis and interpretation | https://data.humdata.org/dataset/kazakhstan-coronavirus-covid-19-subnational-cases. [229] | [229] |
| OCHA Philippines | COVID-19-operational presence risk communication and community engagement in the Philippines | Risk communication and community engagement | https://data.humdata.org/dataset/philippines-covid-19-operational-presence-risk-communication-and-community-engagement-rcce. [230] | [230] |
| Hub Latin America | Epidemiological and hospital indicators on COVID-19 in Ouro Preto, Brazil | For data analysis and interpretation | https://data.humdata.org/dataset/brazil-epidemiological-and-hospital-indicators-on-covid-19-in-ouro-preto. [231] | [231] |
| Safeture | COVID-19 subnational cases in Oman | Reporting cases at a national level. | https://data.humdata.org/dataset/oman-coronavirus-covid-19-subnational-cases. [232] | [232] |
| Humanitarian Emergency Response Africa | Coronavirus [COVID-19] City level cases for Mauritania | Reporting cases at a national level. | https://data.humdata.org/dataset/mauritania-coronavirus-covid-19-city-level. [233] | [233] |
| UNICEF Data and Analytics [HQ] | Tracking children's situation during COVID-19 | Data analysis and interpretation | https://data.humdata.org/dataset/rapid-situation-tracking-for-covid-19-socioeconomic-impacts. [234] | [234] |
| Humanitarian Emergency Response Africa | COVID-19 recoveries in Africa on a national level | Data analysis and interpretation | https://data.humdata.org/dataset/africa-covid-19-recovered-cases. [235] | [235] |
| Mobile Accord, Inc. [GeoPoll] | COVID-19 vaccines and impacts accepted in Sub-Saharan Africa | Data analysis and interpretation | https://data.humdata.org/dataset/covid19-impacts-and-vaccine-acceptance-in-sub-saharan-africa. [236] | [236] |
| United Nations Development Coordination Office | UN Collective Results on the COVID-19 Socioeconomic Response in 2020 | Monitor the progress and achievements of UNCT's collective action in socio-economic response. | https://data.humdata.org/dataset/un-collective-results-on-the-covid-19-socioeconomic-response-in-2020. [237] | [237] |
| Mobile Accord, Inc. [GeoPoll] | Economic impact of COVID-19 in Sub Saharan Africa | Data interpretation and analysis | https://data.humdata.org/dataset/economic-impact-of-covid-19-in-sub-saharan-africa. [238] | [238] |
| HDX | COVID-19 subnational cases in Myanmar | Reporting cases at a national level. | https://data.humdata.org/dataset/myanmar-coronavirus-covid-19-subnational-cases. [239] | [239] |
| Safeture | COVID-19 sub-national cases in Ghana | Reporting cases at a national level. | https://data.humdata.org/dataset/ghana-coronavirus-covid-19-subnational-cases [240] | [240] |
| Insecurity Insight | Covid 19 and Aid Security | To help aid agencies meet the duty of care obligations to staff and reach people in need. | https://data.humdata.org/dataset/aid-security-and-covid-19. [241] | [241] |
| Infoculture | Registry of Russian NGO's affected by COVID-19 | For data analysis and interpretation. | https://data.humdata.org/dataset/ngos-affected-by-covid19-russia. [242] | [242][243] |

| HDX | Facility Interim Distribution Forecast for Covax | For data analysis and interpretation. | https://data.humdata.org/dataset/covax-facility-interim-distribution-forecast. [244] | [244] |
|---|---|---|---|---|
| UNHCR - The UN Refugee Agency | Socio-economic impact of COVID-19 on refugees in Kenya | For data analysis and interpretation. | https://data.humdata.org/dataset/unhcr-ken-2020-socioeconomic-impact-of-covid-19-on-pocs-in-kenya-v2-2. [245] | [245] |
| Humanitarian Emergency Response Africa | COVID-19 city level cases in Togo | Reporting cases at a national level. | https://data.humdata.org/dataset/togo-coronavirus-covid-19-city-level. [246] | [246] |
| UNICEF Data and Analytics | Indicators of interest to COVID-19 data at UNICEF | For data analysis and interpretation. | https://data.humdata.org/dataset/unicef-indicators-of-interest-to-the-covid-19-outbreak. [247] | [247] |
| HDX | COVID-19 subnational cases in Mozambique | Reporting cases at a national level. | https://data.humdata.org/dataset/mozambique-coronavirus-covid-19-subnational-cases. [248] | [248] |
| HDX | COVID-19 subnational cases for Haiti | Reporting cases at a national level. | https://data.humdata.org/dataset/haiti-covid-19-subnational-cases [249] | [249] |
| OCHA HQ | COVID-19 Pandemic induced Humanitarian Access Constraints | For data analysis and interpretation. | https://data.humdata.org/dataset/constraints-faced-by-people-due-to-covid-19-outbreak. [250] | [250] |
| OCHA Philippines | 2020 Significant events happening in Philippines | For data analysis and interpretation. | https://data.humdata.org/dataset/philippines-2020-significant-events. [251] | [251] |
| UNHCR - The UN Refugee Agency | Socio-economic impact of COVID-19 on refugees in Kenya round 5 | For data analysis and interpretation. | https://data.humdata.org/dataset/unhcr-ken-2020-covid-round5-v2-1. [252] | [252][253] |
| OCHA Sudan | COVID-19 response and preparedness 4W in Sudan | COVID-19 response outcomes | https://data.humdata.org/dataset/sudan-covid-19-preparedness-and-response-4w. [254] | [254] |
| Indonesian Red Cross [PMI] | Community Feedback by Indonesian Red Cross [PMI] | COVID-19 response outcomes | https://data.humdata.org/dataset/community-feedback-by-indonesian-red-cross-pmi. [255] | [255] |
| Johns Hopkins Applied Physics Lab | Projected COVID-19 sub-national cases in Sudan | For data analysis and interpretation | https://data.humdata.org/dataset/sudan-projected-covid-19-sub-national-cases. [256] | [256] |
| International Organization for Migration | IATA travel restriction monitoring | For data analysis and interpretation. | https://data.humdata.org/dataset/travel-restriction-monitoring-iata-covid-19-iom-dtm. [257] | [257] |
| OCHA ROWCA | COVID-19 situation in West and Central Africa | For data analysis and interpretation. | https://data.humdata.org/dataset/west-and-central-africa-coronavirus-covid-19-situation. [258] | [258] |
| Johns Hopkins Applied Physics Lab | Projected COVID-19 sub-national cases for Somalia | For data analysis and interpretation. | https://data.humdata.org/dataset/somalia-projected-covid-19-sub-national-cases [259] | [259] |
| OCHA Ethiopia | COVID-19 sub-national cases for Ethiopia. | Reporting cases at a national level. | https://data.humdata.org/dataset/ethiopia-coronavirus-covid-19-subnational-cases [260] | [260] |
| Infoculture | COVID-19 cases in Russia | Reporting cases at a national level. | https://data.humdata.org/dataset/covid-19-cases-data-in-russia. [261] | [261] |
| Mobile Accord, Inc. [GeoPoll] | Ongoing impacts of COVID-19 in Sub-Saharan Africa | For data analysis and interpretation. | https://data.humdata.org/dataset/ongoing-impacts-of-covid-19-in-sub-saharan-africa. [262] | [262] |
| OCHA HQ | Global appeals and plans of COVID-19 around the globe | For data analysis and interpretation. | https://data.humdata.org/dataset/covid-19-global-appeals-and-plans. [263] | [263] |
| Mobile Accord, Inc. [GeoPoll] | Community perception and knowledge of Covid 19 in sub-Saharan Africa | For data analysis and interpretation. | https://data.humdata.org/dataset/coronavirus-in-sub-saharan-africa [264] | [264] |
| OCHA HQ | COVID-19 allocations for CERF and CBPF | Monitoring resource allocation | https://data.humdata.org/dataset/cerf-covid-19-allocations. [265] | [265] |
| INFORM | INFORM COVID-19 comparability and analysis tool | Identification of countries at risk from health and | https://data.humdata.org/dataset/inform-covid-analysis-v01.  [266] | [266] |

| | | humanitarian impacts of COVID-19 that could overwhelm current national response capacity, and therefore lead to a need for additional international assistance | | |
|---|---|---|---|---|
| OCHA HQ | Covid 19 impacts, mitigation, and humanitarian access constraint. | For data analysis and interpretation. | https://data.humdata.org/dataset/covid19-humanitarian-access. [267] | [267] |
| HDX | LSHTM COVID-19 Projections. | For data analysis and interpretation. | https://data.humdata.org/dataset/lshtm-covid-19-projections [268] | [268] |
| UNICEF ESARO | UNICEF COVID-19 response and situation in Eastern and Southern Africa | For data analysis and interpretation. | https://data.humdata.org/dataset/eastern-and-southern-africa-covid-19-unicef-situation-and-response [269] | [269] |
| OCHA Mali | COVID-19 subnational cases in Mali | Reporting cases at a national level. | https://data.humdata.org/dataset/mali-coronavirus-covid-19-subnational-cases. [270] | [270] |
| OCHA HQ | Economic exposure index for COVID-19 | For data analysis and interpretation. | https://data.humdata.org/dataset/covid-19-economic-exposure-index. [271] | [271] |
| OCHA Somalia | COVID-19 sub-national cases for Somalia | Reporting cases at a national level. | https://data.humdata.org/dataset/somalia-coronavirus-covid-19-subnational-cases. [272] | [272][273] |
| Johns Hopkins Applied Physics Lab | Projected COVID-19 sub-national cases for Afghanistan | Reporting cases at a national level. | https://data.humdata.org/dataset/afghanistan-projected-covid-19-sub-national-cases. [275] | [274][275] |
| UNHCR - The UN Refugee Agency | Testing, knowledge, and mask-wearing | For data analysis and interpretation. | https://data.humdata.org/dataset/unhcr-bgd-2020-covid-mwtk-v2-1. [276] | [276] |
| Uganda Red Cross Society | COVID-19 risk index | For data analysis and interpretation. | https://data.humdata.org/dataset/covid19_risk_index-zip. [277] | [277] |
| HDX | COVID-19 subnational cases for the Democratic Republic of Congo | Reporting cases at a national level. | https://data.humdata.org/dataset/democratic-republic-of-the-congo-coronavirus-covid-19-subnational-cases. [278] | [278] |
| HDX | COVID-19 subnational data for Libya | Reporting cases at a national level. | https://data.humdata.org/dataset/libya-coronavirus-covid-19-subnational-cases. [279] | [279] |
| UNHCR - The UN Refugee Agency | Round 2 Socio-economic impacts of COVID-19 on refugees in Kenya | For data analysis and interpretation. | https://data.humdata.org/dataset/unhcr-ken-2020-socioeconomic-impact-of-covid-19-on-pocs-in-kenya-round2-v1-0. [280] | [280] |
| International Organization for Migration | Cameroon COVID-19 Mobility Restriction - Point of Entries - [IOM DTM] | For data analysis and interpretation. | https://data.humdata.org/dataset/cameroon-covid-19-mobility-restriction-point-of-entries-iom-dtm. [281] | [281] |
| UNHCR - The UN Refugee Agency | A panel study of the socio-economic impacts of COVID-19 on refugees living in Kenya | For data analysis and interpretation. | https://data.humdata.org/dataset/unhcr-ken-2020-covid-panel-v2-1. [282] | [282] |
| Johns Hopkins Applied Physics Lab | Projected COVID-19 sub-national cases for Iraq | Reporting cases at a national level. | https://data.humdata.org/dataset/iraq-projected-covid-19-sub-national-cases. [282] | [283] |
| UNHCR - The UN Refugee Agency | Assessment of COVID-19 socio-economic impacts on Persons of concern to UNHCR | Reporting cases at national level. | https://data.humdata.org/dataset/unhcr-nga-2020-sea-covid19-v2-1. [284] | [284][285] |
| UNHCR - The UN Refugee Agency | Assessment of COVID-19 impact on livelihoods of refugees in Zambia | For data analysis and interpretation. | https://data.humdata.org/dataset/ddi-zam-unhcr-covid19-impact-assessment-on-refugee-livelihoods-zambia-july-2020. [286] | [286] |
| Hub Latin America | symptomatology related to the coronavirus COVID-19 in Ecuador | Data exploration, analysis | https://data.humdata.org/dataset/symptomatology-ecu911-santa-cruz-monthly-2018-2021. [287] | [287][288] |
| HDX | Health facilities by province in Afghanistan | For data analysis and interpretation. | https://data.humdata.org/dataset/afghanistan-covid-19-health-facilities-by-province. [289] | [289] |
| ACAPS | COVID-19 humanitarian exceptions | For data analysis and interpretation. | https://data.humdata.org/dataset/acaps-covid-19-humanitarian-exemptions-dataset [290] | [290] |
| International Organization for Migration | COVID-19 mobility and preparedness updates in South Sudan. | For data analysis and interpretation. | https://data.humdata.org/dataset/south-sudan-covid-19-mobility-and-preparedness-updates-iom-dtm. [291] | [291] |
| UNHCR - The UN Refugee Agency | Socio-economic impacts of COVID-19 on refugees living in Kenya, Round 1 | For data analysis and interpretation. | https://data.humdata.org/dataset/unhcr-ken-2020-covid-round1-v2-2. [292] | [292] |
| UNHCR - The UN Refugee Agency | Socio-economic impacts of COVID-19 on refugees living in Kenya, Round 4 | For data analysis and interpretation. | https://data.humdata.org/dataset/unhcr-ken-2020-covid-round4-v2-1. [293] | [293] |
| UNHCR - The UN Refugee Agency | Socio-economic impacts of COVID-19 on refugees living in Kenya, Round 3 | For data analysis and interpretation. | https://data.humdata.org/dataset/unhcr-ken-2020-covid-round3-v2-1. [294] | [294] |
| European Centre for Disease Prevention and Control | COVID-19 vaccination in the EU/EEA | Vaccine administration updates | https://www.ecdc.europa.eu/en/publications-data/data-COVID-19-vaccination-eu-eea. [295] | [295] |
| | Data on the daily number of new reported COVID-19 cases and deaths by EU/EEA country | Monitoring daily COVID-19 cases. | https://www.ecdc.europa.eu/en/publications-data/data-daily-new-cases-COVID-19-eueea-country. [296] | [296] |
| | Data on SARS-CoV-2 variants in the EU/EEA | Monitoring SARS-CoV-2 variants in the EU/EEA | https://www.ecdc.europa.eu/en/publications-data/data-virus-variants-COVID-19-eueea. [297] | [297] |
| | Data on 14-day notification rate of new COVID-19 cases and deaths | Monitoring and analysis of Data on 14-day notification rate of new COVID-19 cases and deaths | https://www.ecdc.europa.eu/en/publications-data/data-national-14-day-notification-rate-COVID-19. [298] | [298][299] |
| | Data on the daily subnational 14-day notification rate of new COVID-19 cases | Monitoring and analysis. | https://www.ecdc.europa.eu/en/publications-data/subnational-14-day-notification-rate-COVID-19. [300] | [300][301] |
| | Data on hospital and ICU admission rates and current occupancy for COVID-19 | Monitoring and analysis. | https://www.ecdc.europa.eu/en/publications-data/download-data-hospital-and-icu-admission-rates-and-current-occupancy-COVID-19. [302] | [302][303] |
| | Data on country response measures | Monitoring and analysis. | https://www.ecdc.europa.eu/en/publications-data/download-data-response-measures-COVID-19 [304] | [304] |
| | Data on age-specific notification rate | Monitoring and analysis. | https://www.ecdc.europa.eu/en/publications-data/COVID-19-data-14-day-age-notification-rate-new-cases. [305] | [305] |
| | Data on council recommendations for mapping the coordinated approach to the restriction of free movement in response to the COVID-19 pandemic in the EU/EEA | Monitoring and analysis. | https://www.ecdc.europa.eu/en/publications-data/indicators-maps-support-council-recommendation. [306] | [306] |
| | Historical data on the COVID-19 daily number of cases and deaths by country, worldwide | Monitoring and analysis. | https://www.ecdc.europa.eu/en/publications-data/download-todays-data-geographic-distribution-COVID-19-cases-worldwide. [307] | [307] |
| Kaggle.com | Daily information on the number of COVID-19 | Monitoring and analysis. | https://www.kaggle.com/sudalairajkumar/novel-corona-virus-2019-dataset [308] | [308] |

| | affected areas across the globe | | | |
|---|---|---|---|---|
| World Health Organization | Information on country reported public measures to curb COVID-19. | Monitoring and analysis | https://www.who.int/emergencies/diseases/novel-coronavirus-2019/phsm. [309] | [309] |
| Johns Hopkins' electronic medical record, Epic | Information on the patients that have been confirmed or are suspected of having contracted COVID-19 | For retrospective analysis of COVID-19 patient populations | https://ictr.johnshopkins.edu/coronavirus/jh-crown/ [310] | [310] |
| National Patient-Centered Clinical Research Network | Data model tracking insights on patients infected with COVID-19 | For understanding and defining demographics infected with SARS-CoV-2 | https://pcornet.org/news/pcornet-COVID-19-common-data-model-launched-enabling-rapid-capture-of-insights/[311] | [311] |
| Johns Hopkins COVID-19 collaboration platform | Publicizing protocols whose PIs are open to various levels of collaboration. | Protocol collaboration. | https://covidcp.org/. [312] | [312] |
| National COVID Cohort Collaborative | Building a centralized national data resource that the research community can use to study COVID-19 and identify potential treatments as the pandemic continues to evolve. | Rapid collection and analysis of clinical, laboratory, and diagnostic data from hospitals and health care plans | https://ncats.nih.gov/n3c/about. [313] | [313] |
| 4CE | COVID-19 positive cases and new death rates by country, overtime | For data analysis and interpretation | https://covidclinical.net/plots/paper-01/release-2020-04-11/dailycounts.html. [314] | [314] |
| 4CE | COVID-19 number of patients by country, by gender | For data analysis and interpretation | https://covidclinical.net/plots/paper-01/release-2020-04-11/demographics.html [315] | [315] |
| 4CE | COVID-19 lab values corresponding to 14 LOINC Codes | For data analysis and interpretation | https://covidclinical.net/plots/paper-01/release-2020-04-11/labs.html [316] | [316] |
| 4CE | Comparison of data from CSSE JHU | For data analysis and interpretation | https://covidclinical.net/plots/paper-01/release-2020-04-11/change.html. [317] | [317] |
| 4CE | Participating sites visualized on maps | For data analysis and interpretation | https://covidclinical.net/plots/paper-01/release-2020-04-11/sites.html [318] | [318] |
| 4CE | Daily Count Data for International Electronic Health Record-Derived COVID-19 Clinical Course Profile | For data analysis and interpretation | https://figshare.com/articles/dataset/Daily_Count_Data_for_International_Electronic_Health_Record-Derived_COVID-19_Clinical_Course_Profile_The_4CE_Consortium/12152976/1. [319] | [319] |
| 4CE | Demographic data for International Electronic Health Record-Derived COVID-19 Clinical Course Profile. | For data analysis and interpretation | https://figshare.com/articles/dataset/Demographics_Data_for_International_Electronic_Health_Record-Derived_COVID-19_Clinical_Course_Profile_The_4CE_Consortium/12152973/1 [320] | [320] |
| 4CE | Diagnosis data for International Electronic Health Record-Derived COVID-19 Clinical Course Profile. | For data analysis and interpretation | https://figshare.com/articles/dataset/Diagnosis_Data_for_International_Electronic_Health_Record-Derived_COVID-19_Clinical_Course_Profile_The_4CE_Consortium/12152967 [321] | [321] |
| 4CE | Labs data for International Electronic Health Record-Derived COVID-19 Clinical Course Profile. | For data analysis and interpretation | https://figshare.com/articles/dataset/Labs_Data_for_International_Electronic_Health_Record-Derived_COVID-19_Clinical_Course_Profile_The_4CE_Consortium/12152766 [322] | [322] |
| 4CE | Labs data for International Electronic Health Record-Derived COVID-19 | For data analysis and interpretation | https://figshare.com/articles/dataset/Healthcare_Systems/12118911 [323] | [323] |

| | Clinical Course Profile. | | | |
|---|---|---|---|---|
| 4CE | Time series COVID-19 confirmed cases | For data analysis and interpretation | https://github.com/CSSEGISandData/COVID-19/blob/dcd4181613f512a6f75249fc77b63286aebe7271/csse_covid_19_data/csse_covid_19_time_series/time_series_covid19_confirmed_global.csv [324] | [324] |
| Health and Retirement Study | 2020 HRS COVID-19 project | For data analysis and interpretation | https://hrsdata.isr.umich.edu/data-products/2020-hrs-COVID-19-project. [325] | [325] |
| COVID-19 research database | Electronic health records, claims, and consumer data. | For data analysis and interpretation | https://covid19researchdatabase.org/. [326] | [326] |
| COVID-19 Research Initiatives in the HRS International Network | HRS COVID-19 Data on questionnaires, surveys, interviews, and state policies | For data analysis and interpretation | https://hrs.isr.umich.edu/data-products/COVID-19 [327] | [327] |
| Center for Disease Control and Prevention | COVID-19 Case Surveillance Public Use Data with Geography | For analysis and interpretation | https://data.cdc.gov/Case-Surveillance/COVID-19-Case-Surveillance-Public-Use-Data-with-Ge/n8mc-b4w4. [328] | [328] |
| Center for Disease Control and Prevention | COVID-19 Case Surveillance Public Use Data | For analysis and interpretation | https://data.cdc.gov/Case-Surveillance/COVID-19-Case-Surveillance-Public-Use-Data/vbim-akqf. [329] | [329] |
| Center for Disease Control and Prevention | COVID-19 Case Surveillance Restricted Access Detailed Data | For analysis and interpretation | https://data.cdc.gov/Case-Surveillance/COVID-19-Case-Surveillance-Restricted-Access-Detai/mbd7-r32t. [330] | [330] |
| Center for Disease Control and Prevention | COVID-19 Vaccine Distribution Allocations by Jurisdiction – Janssen | For analysis and interpretation | https://data.cdc.gov/Vaccinations/COVID-19-Vaccine-Distribution-Allocations-by-Juris/w9zu-fywh [331] | [331] |
| Center for Disease Control and Prevention | COVID-19 Vaccine Distribution Allocations by Jurisdiction - Pfizer | For analysis and interpretation | https://data.cdc.gov/Vaccinations/COVID-19-Vaccine-Distribution-Allocations-by-Juris/saz5-9hgg [332] | [332][333] |
| Center for Disease Control and Prevention | United States COVID-19 Cases and Deaths by State over Time | For analysis and interpretation | https://data.cdc.gov/Case-Surveillance/United-States-COVID-19-Cases-and-Deaths-by-State-o/9mfq-cb36 [334] | [334] |
| Center for Disease Control and Prevention | COVID-19 Vaccine Distribution Allocations by Jurisdiction – Moderna | For analysis and interpretation | https://data.cdc.gov/Vaccinations/COVID-19-Vaccine-Distribution-Allocations-by-Juris/b7pe-5nws. [335] | [335][336] |
| Center for Disease Control and Prevention | Provider Relief Fund COVID-19 Nursing Home Quality Incentive Program | For analysis and interpretation | https://data.cdc.gov/Administrative/Provider-Relief-Fund-COVID-19-Nursing-Home-Quality/bfqg-cb6d [337] | [337] |
| Center for Disease Control and Prevention | Indicators of Anxiety or Depression Based on Reported Frequency of Symptoms During Last 7 Days | For analysis and interpretation | https://data.cdc.gov/NCHS/Indicators-of-Anxiety-or-Depression-Based-on-Repor/8pt5-q6wp [338] | [338] |
| Center for Disease Control and Prevention | Mental Health Care in the Last 4 Weeks | For analysis and interpretation | https://data.cdc.gov/NCHS/Mental-Health-Care-in-the-Last-4-Weeks/yni7-er2q [339] | [339][340] |
| Center for Disease Control and Prevention | Vaccine Hesitancy for COVID-19: County and local estimate | For analysis and interpretation | https://data.cdc.gov/Vaccinations/Vaccine-Hesitancy-for-COVID-19-County-and-local-es/q9mh-h2tw. [341] | [341] |
| Center for Disease Control and Prevention | Loss of Work Due to Illness from COVID-19 | For analysis and interpretation | https://data.cdc.gov/NCHS/Loss-of-Work-Due-to-Illness-from-COVID-19/qgkx-mswu. [342] | [342] |
| Center for Disease Control and Prevention | COVID-19 Vaccinations in the United States by Jurisdiction | For analysis and interpretation | https://data.cdc.gov/Vaccinations/COVID-19-Vaccinations-in-the-United-States-Jurisdi/unsk-b7fc. [343] | [343] |
| Center for Disease Control and Prevention | Provider Relief Fund & Accelerated and Advance Payments | For analysis and interpretation | https://data.cdc.gov/Administrative/Provider-Relief-Fund-Accelerated-and-Advance-Payme/v2pi-w3up [344] | [344] |
| Center for Disease Control and Prevention | Indicators of Reduced Access to Care Due to the Coronavirus Pandemic During Last 4 Weeks | For analysis and interpretation | https://data.cdc.gov/NCHS/Indicators-of-Reduced-Access-to-Care-Due-to-the-Co/xb3p-q62w. [345] | [345] |

| Center for Disease Control and Prevention | Access and Use of Telemedicine During COVID-19 | For analysis and interpretation | https://data.cdc.gov/NCHS/Access-and-Use-of-Telemedicine-During-COVID-19/8xy9-ubqz. [346] | [346][347] |
| Center for Disease Control and Prevention | COVID-19 Vaccination Trends in the United States, National and Jurisdictional data | For analysis and interpretation | https://data.cdc.gov/Vaccinations/COVID-19-Vaccination-Trends-in-the-United-States-N/rh2h-3yt2 [348] | [348] |
| Center for Disease Control and Prevention | Reduced Access to Care During COVID-19 | For analysis and interpretation | https://data.cdc.gov/NCHS/Reduced-Access-to-Care-During-COVID-19/th9n-ghnr. [349] | [349] |
| Center for Disease Control and Prevention | Telemedicine Use in the Last 4 Weeks | For analysis and interpretation | https://data.cdc.gov/NCHS/Telemedicine-Use-in-the-Last-4-Weeks/h7xa-837u [350] | [350][351] |
| Center for Disease Control and Prevention | Provisional COVID-19 Death Counts in the United States by County | For analysis and interpretation | https://data.cdc.gov/NCHS/Provisional-COVID-19-Death-Counts-in-the-United-St/kn79-hsxy [352] | [352] |
| Center for Disease Control and Prevention | Provisional COVID-19 Deaths: Focus on Ages 0-18 Years | For analysis and interpretation | https://data.cdc.gov/NCHS/Provisional-COVID-19-Deaths-Focus-on-Ages-0-18-Yea/nr4s-juj3 [353] | [353] |
| Center for Disease Control and Prevention | COVID-19 Vaccination and Case Trends by Age Group, United States | For analysis and interpretation | https://data.cdc.gov/Vaccinations/COVID-19-Vaccination-and-Case-Trends-by-Age-Group-/gxj9-t96f. [354] | [354] |
| Center for Disease Control and Prevention | Excess Deaths Associated with COVID-19 | For analysis and interpretation | https://data.cdc.gov/NCHS/Excess-Deaths-Associated-with-COVID-19/xkkf-xrst [355] | [355] |
| Center for Disease Control and Prevention | Indicators of Health Insurance Coverage at the Time of Interview | For analysis and interpretation | https://data.cdc.gov/NCHS/Indicators-of-Health-Insurance-Coverage-at-the-Tim/jb9g-gnvr. [356] | [356][357] |
| Center for Disease Control and Prevention | Provisional COVID-19 Death Counts by Week Ending Date and State | For analysis and interpretation | https://data.cdc.gov/NCHS/Provisional-COVID-19-Death-Counts-by-Week-Ending-D/r8kw-7aab [359] | [358][359] |
| Center for Disease Control and Prevention | COVID-19 Vaccination Demographics in the United States, National data | For analysis and interpretation | https://data.cdc.gov/Vaccinations/COVID-19-Vaccination-Demographics-in-the-United-St/km4m-vcsb [360] | [360] |
| Center for Disease Control and Prevention | Nationwide Survey on Commercial Laboratory Seroprevalence | For analysis and interpretation | https://data.cdc.gov/Laboratory-Surveillance/Nationwide-Commercial-Laboratory-Seroprevalence-Su/d2tw-32xv [361] | [361] |
| Center for Disease Control and Prevention | Survey on COVID-19 Hospital Data from the National Hospital Care | For analysis and interpretation | https://data.cdc.gov/NCHS/COVID-19-Hospital-Data-from-the-National-Hospital-/q3t8-zr7t [362] | [362][363] |
| Center for Disease Control and Prevention | Provisional COVID-19 Death Counts by Age in Years, 2020-2021 | For analysis and interpretation | https://data.cdc.gov/NCHS/Provisional-COVID-19-Death-Counts-by-Age-in-Years-/3apk-4u4f [364] | [364] |
| Center for Disease Control and Prevention | Long-term Care and COVID-19 | For analysis and interpretation | https://data.cdc.gov/NCHS/Long-term-Care-and-COVID-19/3j26-kg6d [365] | [365] |
| Center for Disease Control and Prevention | Provisional COVID-19 Deaths by Place of Death and State | For analysis and interpretation | https://data.cdc.gov/NCHS/Provisional-COVID-19-Deaths-by-Place-of-Death-and-/uggs-hy5q. [366] | [366] |
| Center for Disease Control and Prevention | Provisional COVID-19 Deaths by Week and Urbanicity | For analysis and interpretation | https://data.cdc.gov/NCHS/Provisional-COVID-19-Deaths-by-Week-and-Urbanicity/hkhc-f7hg. [367] | [367][368] |
| Center for Disease Control and Prevention | U.S. State and Territorial Stay-At-Home Orders: March 15, 2020 – August 15, 2021 by County by Day | For analysis and interpretation | https://data.cdc.gov/Policy-Surveillance/U-S-State-and-Territorial-Stay-At-Home-Orders-Marc/y2iy-8irm. [369] | [369] |
| Center for Disease Control and Prevention | U.S. State and Territorial Public Mask Mandates from April 10, 2020 through August 15, 2021 by County by Day | For analysis and interpretation | https://data.cdc.gov/Policy-Surveillance/U-S-State-and-Territorial-Public-Mask-Mandates-Fro/62d6-pm5i [370] | [370][371] |

| Center for Disease Control and Prevention | U.S. State, Territorial, and County Stay-At-Home Orders: March 15-May 5 by County by Day | For analysis and interpretation | https://data.cdc.gov/Policy-Surveillance/U-S-State-Territorial-and-County-Stay-At-Home-Orde/qz3x-mf9n. [372] | [372] |
|---|---|---|---|---|
| NCHS | Provisional Death Counts for Influenza, Pneumonia, and COVID-19 | For analysis and interpretation | https://data.cdc.gov/NCHS/Provisional-Death-Counts-for-Influenza-Pneumonia-a/ynw2-4viq. [373] | [373][374] |
| European COVID-19 data platform | Three data hubs reporting SARS-CoV-2, COVID-19, and Federated European Genome-phenome | For data exploration, analysis, and interpretation. | https://www.covid19dataportal.org/the-european-COVID-19-data-platform [375] | [375] |
| Open Safely | Computational resources and open access data to address COVID-19 | For data exploration, analysis, and interpretation. | https://datascience.nih.gov/COVID-19-open-access-resources [376] | [376] |
| ImmPort Shared Data | Research data available to the public and mostly scientific community to improve research work around COVID-19 | For data exploration, analysis, and interpretation. | https://www.immport.org/shared/search?filters=study_2_condition_or_disease.condition_preferred:COVID-19%20-%20DOID:0080600&utm_source=COVID-19&utm_medium=banner&utm_campaign=COVID-19 [377] | [377] |
| World Health Organization | Global COVID-19 situation for confirmed cases. | For data exploration, analysis, and interpretation. | https://covid19.who.int/. [378] | [378] |
| World meter | Global COVID-19 cases including confirmed cases, deaths, active cases, and closed cases. | For data exploration, analysis, and interpretation. | https://www.worldometers.info/coronavirus/ [379] | [379] |
| The World Bank | COVID-19 household monitoring dashboard. | For data exploration, analysis, and interpretation. | https://www.worldbank.org/en/data/interactive/2020/11/1/COVID-19-high-frequency-monitoring-dashboard [380] | [380] |
| The World Bank Group | COVID-19 business pulse survey dashboard that contains data on the socio-economic impacts of COVID-19 in 76 selected countries. | For data exploration, analysis, and interpretation. | https://www.worldbank.org/en/data/interactive/2021/01/19/COVID-19-business-pulse-survey-dashboard. [381] | [381][382] |
| The World Bank Group | Guidance to World Bank Group vendors on COVID-19. | For data exploration, analysis, and interpretation. | https://www.worldbank.org/en/about/corporate-procurement/announcements/guidance_on_COVID-19 [383] | [383] |
| The World Bank Group | Harmonized COVID-19 household monitoring survey | For data exploration, analysis, and interpretation. | https://datacatalog.worldbank.org/search/dataset/0037769/Harmonized-COVID-19-Household-Monitoring-Surveys. [384] | [384] |
| Centers for Disease Control and Prevention | Effectiveness of COVID-19 vaccines | For data exploration, analysis, and interpretation. | https://www.cdc.gov/coronavirus/2019-ncov/vaccines/effectiveness.html. [385] | [385] |
| Centers for Disease Control and Prevention | COVID-19 integrated country view | For data exploration, analysis, and interpretation. | https://covid.cdc.gov/covid-data-tracker/#county-view. [386] | [386] |
| Centers for Disease Control and Prevention | Forecasting cases and deaths COVID-19 in the United Sates | For data exploration, analysis, and interpretation. | https://covid.cdc.gov/covid-data-tracker/#forecasting_weeklydeaths. [387] | [387] |
| Centers for Disease Control and Prevention | COVID-19 vaccinations in the U.S | For data exploration, analysis, and interpretation. | https://covid.cdc.gov/covid-data-tracker/#vaccinations_vacc-total-admin-rate-total [388] | [388] |
| Centers for Disease Control and Prevention | Country-level vulnerability index in the United States | For data exploration, analysis, and interpretation. | https://covid.cdc.gov/covid-data-tracker/#pandemic-vulnerability-index [389] | [389] |
| Centers for Disease Control and Prevention | COVID-19 community profile report | For data exploration, analysis, and interpretation. | https://healthdata.gov/Health/COVID-19-Community-Profile-Report/gqxm-d9w9 [390] | [390] |

## II. CONCLUSION

The summary tables (Tables I and II) present the technological resources and datasets used in tackling covid-19. Most of the data collected with COVID-19 related to hospitalizations, vaccinations, government response measures, deaths, confirmed reported cases, as well as restrictions and policies are used in aiding the pandemic. The R resources have mainly been used to develop Shiny apps and dashboards. Java, Kotlin, and Perl resources have been used in developing Android and iOS applications for contact tracing, disease surveillance, fast diagnostics, and notifying users anonymously if they have had any contact with someone who has been infected with COVID-19 via low-power bluetooth technology [28][31][32][51][52][69][129]. Based on the benefits of utilizing these resources, continued research and application of technological resources are highly recommendable [70].